# Luminescence centers in proton irradiated single crystal CVD diamond

C. Manfredotti[1,2]*, S. Calusi[2,3], A. Lo Giudice[1,2], L. Giuntini[3],

M. Massi[3], P. Olivero[1,2], A. Re[1,2]

[1] Experimental Physics Department and "Nanostructured Interfaces and Surfaces" Centre of Excellence, University of Torino, via P. Giuria 1, 10125 Torino, Italy

[2] INFN Sezione di Torino, via P. Giuria 1, 10125 Torino, Italy

[3] Physics Department of University and INFN, Firenze, via Sansone 1, 50019 Sesto Fiorentino (Firenze), Italy

* corresponding author (manfredotti@to.infn.it)

**Abstract**

Diamond displays a large variety of luminescence centers which define its optical properties and can be either created or modified by irradiation. The main purpose of the present work is to study the radiation hardness of several of such centers in homoepitaxial single-crystal CVD diamond by following the evolution of photoluminescence and ionoluminescence upon 2 MeV proton irradiation. Luminescence decays were observed with values of the fluence at half of the starting luminescence ($F_{1/2}$) of the order of $10^{14}$ cm$^{-2}$. The 3H center displayed a non-monotonic behavior, with a growing behavior and a subsequent decay with a rather high $F_{1/2}$ value (in the order of few $10^{16}$ cm$^{-2}$), maintaining at the highest fluences an intensity significantly higher than the blue A-band. A simple model based on a double-exponential trend was defined to fit with satisfactory accuracy the evolution of the 3H center. Several PL centers (namely: 3H, TR12, 491 nm, 494 nm) exhibited clear correlations and anti-correlations in their fluence dependences, which were considered in the attempt to acquire some insight into their possible alternative attributions.

**1. Introduction**

Artificial and natural diamonds, in either polycrystalline or monocrystalline form, display a vast variety of optical centers [1], most of which display specific sensitivity to radiation-induced structural damage (induced by electrons, ions, etc.). This implies that such centers are related to specific structural defects in the diamond lattice, and thus can be not only quenched, but also created by radiation itself.

The "radiation hardness" is an important parameter characterizing different luminescence centers, both for fundamental research on the dynamics of defects formation in diamond, and for several applications, such as the employment of these centers as scintillators in radiation detectors [2], or for applications in photonics (i.e. single photon emitters) in which radiation-induced damage is employed as a tool to control their formation [3].

Ionoluminescence (IL) is a powerful and direct method that allows the monitoring in real time of the quenching of intrinsic luminescence during irradiation, as well as the investigation of the more complex evolution of luminescence centers created by irradiation itself [2, 4, 5]. Moreover, IL has the advantage of both inducing structural changes in the material and exciting the related luminescence centers from very specific regions of the sample, since both the nuclear and the electronic energy loss (which are respectively responsible for the two processes) are characterized by well defined profiles which can be readily simulated with Monte Carlo codes, such as SRIM [6].

In this work the evolution of optical centers in CVD single crystal diamond upon 2 MeV proton irradiation was monitored in real time by means of IL as a function of radiation fluence (up to $10^{17}$ cm$^{-2}$, corresponding to about $7 \cdot 10^9$ Gy). Moreover, photoluminescence (PL) was performed on regions implanted at increasing fluences, representing a complementary approach in the study of the evolution of luminescent centers upon ion-induced damage.

In order to have more flexibility in the acquisition of extended spectra or in the monitoring of specific emissions, IL was collected using a versatile double apparatus consisting in a CCD detector array allowing the acquisition of spectra in the 200–900 nm spectral range, and in a grating monochromator coupled with a photomultiplier to follow the trend of at a specific wavelength with more sensitivity and fastness.

The IL and PL results are presented and discussed in order to obtain new insight in the study of radiation-related luminescent centers in diamond, particularly with regards to the new generation of homoepitaxial CVD monocrystalline samples that has been developed in recent years.

**2. Experimental**

The samples under analysis are type IIa single-crystal diamonds grown with Chemical Vapor Deposition (CVD) technique by ElementSix with size $3.0 \times 3.0 \times 0.5$ mm$^3$ [7]. The crystals consist of a single {100} growth sector and the concentrations of nitrogen and boron impurities are below 0.1 ppm and 0.05 ppm, respectively.

IL measurements were performed at external scanning micro-beam facility of the 3 MV Tandetron accelerator of the INFN LABEC Laboratory in Firenze [8]. Luminescence was induced by a 2 MeV proton beam extracted in air or in He atmosphere through a silicon nitride window (100 nm thick, 1 mm$^2$ area). The 10 μm proton beam was raster scanned by means of an Oxford Microbeam system over the sample located 2.5 mm down-stream of the exit window. 150×150 μm$^2$ areas were implanted at a fluences in the $10^{13}$-$10^{17}$ cm$^{-2}$ range, with a 0.1-1.2 nA beam current. The IL apparatus was recently developed and consist of two easily interchangeable systems for light collection, optically coupled to the sample through a system of optical fibers system focused on the beam impact point. The two systems consist in an Ocean Optics USB2000 spectrometer with spectral range 200–900 nm (resolution of about 2 nm) and a grating monochromator with 300–800 nm spectral range coupled with a photomultiplier; in the latter case, the spectral resolution is about 4-9 nm depending from the selected wavelength. The spectrometer was used to collect IL spectra in "one shot" over a broad spectral range: the possibility of acquiring in at the same time the signal intensity over the whole spectrum is very relevant in the case of ion luminescence, where rapid damage-related changes in the material could induce significant distortion in a spectrum acquired sequentially, i.e. with a scanning grating. Complementarily, the photomultiplier was employed to monitor with higher sensitivity and fast response time the temporal evolution of IL signal at a fixed wavelength or to acquire maps by means of the synchronization with the ion beam scanning system. The measured spectra were corrected for the spectral response of the instruments, which were suitably calibrated with test lamps. More details on IL set-up can be found in previous works [2, 9]. Moreover, in order to monitor in real time the exact fluence corresponding to each IL acquisition, both spectrometer and photomultiplier were synchronized with a detector counting the characteristic x-ray emitted from the beam exit window. Counts are proportional to the beam current and a preliminary calibration was carried out with a Faraday cup.

Samples were also analyzed after the implantation by means of photoluminescence (PL). In this case the radiation-damage dependence was evaluated by characterizing with the laser micro-beam different regions implanted at increasing fluences. The PL instrument is a "Renishaw InVia Raman Microscope" equipped with a Kimmon He-Cd laser (excitation wavelength $\lambda$=442 nm). A 2400 lines mm$^{-1}$ diffraction grating and a Leica N Plan 5×/0.12 objective were employed to perform measurements. The output power of laser was 90 mW, but the power at the focal spot on the sample surface was reduced to less than 50 µW with the use of an attenuating filter. The acquisition time for PL measurements was 10 s in the extended range 460 – 520 nm.

## 3. Results and discussion

The three-dimensional plot in Fig. 1 displays the evolution of the IL spectrum as a function of fluence in the $1 \cdot 10^{14} – 2 \cdot 10^{15}$ cm$^{-2}$ range. The fast decay of the NV$^0$ emission with zero phonon line (ZPL) at $\lambda$=575 nm and its relative phonon sidebands is clearly visible. The NV$^0$ emission is commonly attributed to the nitrogen-vacancy complex in neutral charge state [10]. On the other hand, the emission at $\lambda$=503 nm exhibits a significant increase in the above-mentioned fluence range. We attribute the latter emission to the 3H center, since it is not associated with the N3 emission at $\lambda$=415 nm, as usual for the H3, which represents a possible alternative attribution [1]. The 3H center is associated with structural damage induced by any type of radiation and its most widespread attribution is based on the formation of the <100> split self-interstitial [11].

A better view of typical IL spectra is shown in Fig. 2, where the first and last of the spectra of another implantation run are reported. In the plot of Fig. 2a (relevant to a fluence of $1 \cdot 10^{14}$ cm$^{-2}$) the 3H center is at the first stages of its formation, while the NV$^0$ zero phonon line and its phonon replica at $\lambda$=587, 601, 619 nm are still very intense. Moreover, a peak at $\lambda$=532 nm is clearly visible, which is frequently observed in CVD diamond films along with the NV$^0$ emission, and tentatively attributed to a nitrogen containing defect (possibly a nitrogen-vacancy complex) [12]. In the plot of Fig. 2b (relevant to a fluence of $4 \cdot 10^{16}$ cm$^{-2}$) the NV$^0$ emission and the 532 nm line have completely disappeared, while the 3H emission (together with its phonon replica at $\lambda$=518 nm) is strongly enhanced. Moreover, other damage-related peaks are visible at $\lambda$=389 nm and $\lambda$=471 nm. Unfortunately the 389 nm diamond IL emission is superimposed with the luminescence from the helium atmosphere surrounding the sample which happens to occur at about the same wavelength; therefore no analysis will be carried on this emission line. For the same reason, the $\lambda$=425 nm line, which is to be entirely attributed to the emission from the helium atmosphere, will not be considered. The emission at $\lambda$=471 nm is attributed to the TR12 center, which is reported to be a typical radiation center observed in all types of diamond after any high-energy implantation [1]; it is tentatively attributed to a defect involving two vacancies and two interstitial carbon atoms in hexagonal positions [12].

As mentioned above, PL measurements were carried out with λ=442 nm laser excitation in the spectral region from 440 nm to 530 nm, on regions implanted at increasing fluences. Fig. 3 shows typical PL spectra collected from a pristine region, as well as from regions implanted at low (F = 9·$10^{14}$ $cm^{-2}$) and high (F = 4·$10^{16}$ $cm^{-2}$) fluence. Also in this case the 3H and TR12 emissions are strong at high damage densities. Moreover, other centers appear only in PL at λ=491 nm and λ=494 nm. The λ=491 nm emission has been documented in previous cathodoluminescence studies from various types of diamond as an unambiguously radiation-related center [13], but its attribution is still unclear [1]. On the other hand, the λ=494 nm emission can be identified to either the H4 center or to the anti-Stokes emission of the 3H center. In the former case, the attribution is related to a typical radiation center observed in many types of diamonds with several techniques (cathodoluminescence, photoluminescence, absorption, etc.) after different kind of irradiations [1]; the most widely used model for the center is the 3N-V-V-N complex [14]. Alternatively, the latter attribution is related to the strong anti-Stokes phonon sideband of the 3H emission at 408 $cm^{-1}$ shift, as reported in [15].

The decay of the IL emission peaks as a function of implantation fluence was systematically measured with both the described light collection systems; the plot in Fig. 4 shows the decays of the IL signal from the $NV^0$, 533 nm, 3H and blue A-band emissions (depending by the analysed peak the band pass was in the range 6-20 nm).

The above mentioned decays were analyzed as follows.

In the case of $NV^0$ center, data were fitted with the Birks-Black model, which has been employed in previous works for investigating damage effects in scintillators [16], as well as in diamond [2, 4, 17].

$$L(F) = \frac{L_0}{\left(1 + \dfrac{F}{F_{1/2}}\right)} \quad (1)$$

where L(F) and $L_0$ are the luminescence intensities (expressed in arbitrary units) at a given fluence F and at zero fluence, respectively.

This parametric model yields the fluence $F_{1/2}$ at which the luminescence is reduced to 50% of its initial value. An excellent consistency of results was obtained for different implantation runs performed on several samples, with a resulting average value of $F_{1/2}=(1.13 \pm 0.07) \cdot 10^{14}$ cm$^{-2}$. Similarly, the fluence dependences of the 533 nm emission and of the blue A-band were fitted with the above-mentioned model. In this case, $F_{1/2}$ values of $(1.1 \pm 0.2) \cdot 10^{14}$ cm$^{-2}$ and $(1,53 \pm 0,15) \cdot 10^{14}$ cm$^{-2}$ were obtained. $F_{1/2}$ values for $NV^0$ and 533 center are consistent and this could confirm the correlation to nitrogen containing defects observed in [1] [12].

The decay of A-band luminescence was investigated in more details with the formula derived by Sullivan and Baragiola in their IL report on the 30 keV C implantation in diamond [17]. In their work, the data were fitted with the following equation:

$$L(F) = \frac{L_0}{1 + k \cdot [\exp(\sigma \cdot F) - 1]} \quad (2)$$

where σ is the effective damage cross section and k is the ratio between the non-radiative and radiative transition rates for the A-band luminescence. By adopting the k=300 value resulting from [17], from the fitting of our experimental data we obtain $\sigma=(2.2\pm0.2)\cdot10^{-17}$ cm$^2$, which is lower by a factor of about 180 with respect to what reported in [17]. Such a significant difference can be qualitatively explained if the different damage profiles of 30 keV C and 2 MeV H ions are considered. We assume that the damage is proportional to the nuclear energy loss of the impinging ions, which can be evaluated for both implantations in the form of a linear density of induced vacancies per single impinging ion with SRIM Monte Carlo simulation code [6] by setting the displacement energy in diamond lattice to 50 eV [18]. According to the simulations, 30 keV C implantation is associated with a maximum damage density of $\sim2.1\cdot10^{-1}$ vacancies Å$^{-1}$ ion$^{-1}$, compared to a value of $\sim4.4\cdot10^{-4}$ vacancies Å$^{-1}$ ion$^{-1}$ for 2 MeV H. The ratio between the two values ($\sim4.8\cdot10^2$) compares with the ratio between the effective cross sections ($\sim1.8\cdot10^2$); the compatibility between the orders of magnitude of the above mentioned values is satisfactory if the drastic simplification of our assumptions is considered.

As shown in Fig. 4d, he 3H center displays a non-monotonic dependence on the implantation fluence since is not only created by ion damage, but it can be quenched by further damage that leads to the formation of other structural configurations at higher defect densities. A simple model was developed to account this fluence behavior, i.e. the initial formation and subsequent destruction of the defects that are responsible for the 3H emission. Similarly to the equations regulating the statistical evolution of populations of radionuclides, the model refers to a double decay process in which the starting species (pristine sp$_3$ sites) is converted by irradiation into its "son" species (3H centers), which by the same irradiation process is converted to a third species (which we identify with any non-luminescent center).

The differential equations describing the variation of the concentrations of the pristine diamond sites ($N_{sp_3}$) and of the 3H sites ($N_{3H}$) can be written as follows:

$$\begin{cases} \dfrac{dN_{sp_3}}{dF} = -\lambda_{sp_3 \to 3H} \cdot N_{sp_3} \\ \dfrac{dN_{3H}}{dF} = -\lambda_{3H \to nr} \cdot N_{3H} - \dfrac{dN_{sp_3}}{dF} \end{cases} \qquad (3)$$

where the $\lambda$ coefficients represent the effective damage cross sections for the conversion from pristine $sp_3$ sites into 3H sites ("$sp_3 \to 3H$" subscript) and from 3H site to generic non-3H defect sites ("$3H \to nr$" subscript).

It is worth noticing that in the second differential equation the variation of the concentration of 3H sites is not only determined by its intrinsic damage-induced decay, but also by the decay of $sp_3$ sites into 3H sites.

The above mentioned equations can be integrated with the following boundary conditions:

$$\begin{cases} N_{sp_3}(F=0) = N^0_{sp_3} \\ N_{3H}(F=0) = 0 \end{cases} \qquad (4)$$

where $N^0_{sp_3}$ is the initial concentration of pristine $sp_3$ sites.

The resulting equations describe the variation of the densities of sp$_3$ and 3H sites as a function of ion fluence:

$$\begin{cases} N_{sp_3}(F) = N^0_{sp_3} \cdot \exp(-\lambda_{sp_3 \to 3H}) \\ N_{3H}(F) = N^0_{sp_3} \cdot \dfrac{\lambda_{sp_3 \to 3H}}{\lambda_{sp_3 \to 3H} - \lambda_{3H \to nr}} \cdot [\exp(-\lambda_{3H \to nr} \cdot F) - \exp(-\lambda_{sp_3 \to 3H} \cdot F)] \end{cases} \quad (5)$$

The IL intensity from 3H centers is directly proportional to their density, and can be expressed as follows:

$$I_{3H}(F) = I^0_{sp_3} \cdot \dfrac{\lambda_{sp_3 \to 3H}}{\lambda_{sp_3 \to 3H} - \lambda_{3H \to nr}} \cdot [\exp(-\lambda_{3H \to nr} \cdot F) - \exp(-\lambda_{sp_3 \to 3H} \cdot F)] \quad (6)$$

Eq. (6) describes the well-known "double exponential" function employed in the statistical theory of radionuclides. In this context, the inverse of the cross sections $\lambda$ take the role that lifetimes constants have in the classical theory (where time is the independent variable), and define the characteristic fluences that modulate the exponential decays of the pristine sp$_3$ sites into 3H, and of 3H sites into non-radiative centers.

As shown in Fig. 4d, the experimental data are well interpolated by the fitting function reported in eq. (6), where the cross sections $\lambda$, as well as the initial luminescence intensity $I^0_{sp_3}$, are varied as fitting parameters. The compatibility between experimental data and the fitting function is particularly striking if the relative simplicity of the adopted model is considered, as compared by the complexity of the damage-induced structural changes under investigation. After the best fitting parameters are determined, the relevant $F_{1/2}$ values can be evaluated with the well-known equation $F_{1/2} = \ln(2) / \lambda$, yielding values of $(3.18 \pm 0.05)\ 10^{15}\ cm^{-2}$ and $(3.42 \pm 0.06)\ 10^{16}\ cm^{-2}$ for the damage-induced creation and annihilation of 3H centers. It is worth noticing that the former characteristic fluence is about one order of magnitude lower than the latter one, implying that the center is readily formed in diamond while being more resilient to damage quenching.

As mentioned above, also PL measurements were carried out over a series of regions irradiated at increasing fluences, in order to verify the agreement between the IL measurements collected in real time from a single region while ion-induced damage is taking effect, and the PL yield measured from different regions after the previous multiple implantations. Indeed, the two types of measurements yield significantly different results: in particular, in the "luminescence intensity vs fluence" plots for the 3H emission it was observed that analogous changes in luminescence signal were occurring in PL at significantly higher fluence values with respect to IL, where analogous effects where taking place more readily, i.e. for smaller fluence variations. In particular, only the initial increase of 3H emission could be observed with PL. This evidence can be explained if it is considered that in IL the emitted light is generated within the entire depth crossed by incoming ions (i.e. ~25 μm for 2 MeV H ions) along the strongly non-uniform damage profile, and particularly from the Bragg's peak at the end of the ion range where most of the electron energy loss is stimulating the luminescence signal, as shown in Fig. 5. On the other hand, in confocal PL the collected light arises from the very first microns below the surface, a region in which nuclear damage is significantly lower.

Therefore, since the regions probed by PL are less damaged than the regions probed by IL, it is reasonable to expect to measure with PL the same damage-induced effects at significantly higher fluences. A suitable re-scaling of the fluences for IL and PL measurements is therefore needed to compare the same effects arising from different probed depths. The determination from first principles of the correct scale factors is indeed rather complex, since the different damage processes at different probed depths would need to be taken into account, together with complex shadowing effects due to the attenuation of the generated luminescence light when crossing damaged regions.

Therefore, the scale factor for the re-scaling of the fluence values in the PL data was in a first instance evaluated empirically. As shown in Fig. 6, if the fluence data relevant to PL measurements are divided by a factor $10^2$, we obtain an excellent agreement between the "luminescence vs fluence" trends relevant to PL and IL. It is worth stressing that also a suitable re-normalization of the luminescence intensities that were measured with different experimental apparatuses. After the $10^2$ scaling factor is empirically determined, it can be used to estimate the thickness of the layer probed in PL confocal measurements if a very simple assumption is made to compare the intensities of the luminescence yields with different excitation depth profiles. In the case of IL, we assume that the yield from 3H centers is proportional to the integral of the strongly non-uniform damage depth profile over the whole ion range (blue curve in Fig. 5) after weighting it with the non-uniform excitation profile determined by ionization events (dashed red curve in Fig. 5). On the other hand, the PL signal is obtained by integrating the damage depth profile from the surface to a given thickness, representing the probed depth of the confocal laser beam (green box in Fig. 5); in such a layer a uniform excitation profile is assumed. It is worth stressing that in this simple model no self-absorption processes from damaged layers are taken into account. If a factor of $10^2$ is adopted from the empirical re-scaling of the fluences, then the resulting thickness probed in PL measurements is estimated to be ~2 μm. Such a value is in very satisfactory agreement with the expected probing depth of the confocal apparatus, despite the above mentioned significant approximations. It is worth remarking that the validity of the above mentioned considerations is based on the assumption that it is possible to compare the luminescence yield from a sample during its implantation (as in IL) and few days after the implantation of different regions (as in PL), i.e. no significant structural modifications take place in the sample at room temperature after the ion implantation is performed.

In conclusion we report in Fig. 7 the "luminescence vs fluence" plots relevant to the above mentioned luminescence peaks measured in PL, namely the TR12 emission at λ=471 nm, the 3H emission at λ=503 nm, as well as the 491 nm and 494 nm peaks. The various centers display significantly different trends in the $10^{15} - 10^{17}$ cm$^{-2}$ fluence range. Firstly, the TR12 emission (Fig. 7a) increases at increasing fluences with no sign of saturation at high damage levels; on the other hand, the other two centers that grow at increasing fluence (3H, 494 nm) display a tendency to saturate towards high fluences (Figs. 7b and 7c). The fact that the 3H and 494 nm emission are characterized by similar trends is confirmed by their strong correlation, as shown in Fig. 7e; this evidence seems to support the attribution of the 494 nm emission to the anti-Stokes phonon sideband of peak 3H [15], although no definitive answer can be given at this stage. The only peak displaying a monotonous decrease in intensity as a function of increasing fluence is the 491 nm emission (Fig. 7d); in this case a strong anti-correlation with respect to the 3H emission is observed (Fig. 7f). Although the attribution of this center is not definitive [13], we can argue that the initial increase in the concentration of <100> split self-interstitial defects (which are responsible for the 3H emission [11]) is correlated with the progressive destruction of the center responsible for the 491 nm emission. Also in this case, no further conclusion can be drawn at this stage.

**4. Conclusions**

IL and PL measurements have been performed on homoepitaxial CVD diamond samples subjected to implantations with 2 MeV H at fluences up to about $\sim 10^{17}$ cm$^{-2}$. The behavior of several luminescence centers (blue A-band, NV$^0$, TR12, 532 nm, 491 nm, 494 nm) was investigated as a function of implantation fluence. The adoption of conventional functional relations, such as the Birks-Black model and the exponential decay formula developed by Sullivan and Baragiola in [17], allowed the estimation of the values of the characteristic fluences $F_{1/2}$ governing the "luminescence vs fluence" variations. A novel "double exponential" model was developed to describe the non-monotonic behavior of the 3H emission, which is reported here for the first time at the best of our knowledge.

Photoluminescence proved to be a complementary technique with respect to ionoluminescence, on the basis of the different probe depths of MeV ion and visible light. Remarkably, IL and PL fluence-dependent data yielded consistent results provided that suitable re-scaling factors were introduced in the respective fluences accounting for the above-mentioned differences in probing depth of the luminescence signal. Finally, the correlation and anti-correlation of specific emissions allowed the acquisition of some insight into the attribution of different centers, although future studies will be needed to provide definitive answers on this subject.

**Acknowledgements**

Thanks are due to Dr. Alessandro Damin for PL measurements with a confocal microRaman apparatus. This work is supported by the "FARE" experiment of "Istituto Nazionale di Fisica Nucleare" (INFN). The work of P. Olivero is supported by the "Accademia Nazionale dei Lincei – Compagnia di San Paolo" Nanotechnology grant, which is gratefully acknowledged.

**Figure captions**

Fig. 1: three-dimensional plot of IL spectra at increasing fluences of 2 MeV H ions in the $1\cdot10^{14} - 2\cdot10^{15}$ cm$^{-2}$ range; the rapid decrease of the NV$^0$ emission at $\lambda$=575 nm is clearly visible, as well as the increase in the 3H emission at $\lambda$=503 nm.

Fig. 2: typical IL spectra taken at the beginning (a) and at the end (b) of an implantation run; the corresponding implantation fluences are F=$1\cdot10^{14}$ cm$^{-2}$ and $4\cdot10^{16}$ cm$^{-2}$, respectively.

Fig. 3: typical PL spectra taken with $\lambda$=442 nm laser excitation from pristine (a) and irradiated (b, c) regions; in the case of irradiated regions, the relevant fluences are F = $9\cdot10^{14}$ cm$^{-2}$ and F = $4\cdot10^{16}$ cm$^{-2}$, respectively. Vertical offset is used.

Fig. 4: evolution of IL yield as a function of implantation fluences as monitored in real time with the "monochromator + photomultiplier" acquisition system (A-band and 3H centers) and USB2000 spectrometer (NV$^0$ and 533 nm center). Fit obtained using Birks-Black model (a) and double-exponential trend (b) are also shown.

Fig. 5 (color online): (blue continuous line) depth profile of the damage linear density expressed in number of vacancies per unit of penetration depth per incoming ion, referred to the left vertical axis; (red dashed line) depth profile of the energy converted in electronic excitation expressed in eV per unit of penetration depth per incoming ion, referred to the left vertical axis; the solid box in green defines the probe depth of the confocal PL setup.

Fig. 6: plot of the luminescence yield versus the implantation fluence measured with IL (red circles) and PL (black squares); the fluences relevant to PL data have been rescaled of a factor $\sim 10^{-2}$, resulting in a satisfactory matching between the different datasets.

Fig. 7: plots of the PL luminescence yield versus the implantation fluence measured for four characteristic emissions: TR12 (a), 3H (b), 494 nm (c) and 491 nm (d); correlation between 3H and 494 nm emissions (e) and between 3H and 491 nm emissions (f).